\documentclass{elsart}

\renewcommand{\thepage}{plb-curves-v17.tex \hfill \arabic{page}}

\usepackage{graphicx}

\usepackage{cite}

\usepackage{amssymb}

\makeatletter
 \def\ps@copyright{\let\@mkboth\@gobbletwo
  \def\@oddhead{}%
  \let\@evenhead\@oddhead
  \def\@oddfoot{\thepage}
  \let\@evenfoot\@oddfoot
} \makeatother

\newcommand{\be}{\begin{eqnarray}}
\newcommand{\ben}{\begin{eqnarray}\nonumber}
\newcommand{\ee}{\end{eqnarray}}

\newcommand{\Int}[2]{\int\limits_{#1}^{#2}}

\begin{document}

\begin{frontmatter}

\title{Capture of cosmic objects by central gravitational field
of a galaxy cluster}

\author[JINR,Kielce]{Piotr
Flin\thanksref{BI}},\ead{sfflin@cyf-kr.edu.pl}
~\author[JINR]{Victor Pervushin},\ead{pervush@thsun1.jinr.ru}
~\author[JINR]{Andrey Zorin}

\address[JINR]{Bogoliubov Laboratory of Theoretical Physics,
Joint Institute for Nuclear Research, 141980, Dubna, Russia}
\address[Kielce]{Pedagogical University, Institute of Physics,
25-406 Kielce, Poland}

\thanks[BI]{Supported by Bogoliubov---Infeld Programme}

\begin{abstract} The effect of capture of a cosmic object by the
central gravitational field of a galaxy cluster is described in
the expanding Universe. The cosmic evolution can be the origin of
the capture explaining formation of galaxies and their clusters
from the homogenous distribution of matter in the Universe. The
Newton--like equation of the capture is derived  for arbitrary
equations of state in terms of the red shift parameter, and the
influence of the Hubble velocities on the rotational curves is
studied. The obtained
 rotational curves   show us that
 the deficit of the visible matter for superclusters $M\sim 10^{15}M_{\odot}$ can be
increased in the version of  cosmology where observational
quantities are identified with the conformal ones.
\end{abstract}

\end{frontmatter}

\section*{Introduction } Mechanism
of formation of galaxies and their clusters from homogenous
distribution of matter in the Universe is not yet understood. The
main problem is to clear up how the energy of two interaction
particles increases to be negative so that the formation of a
bound state is possible. The evolution of the Universe as a
mechanism of braking of a cosmic object in the central
gravitational field was probably firstly considered by  Einstein
and Strauss in 1945 \cite{Einstein-Strauss:45}. This idea was
developed in terms of conformal variables and coordinates in
\cite{astro-ph:0301543,astro-ph:0206114} where it was shown that
the cosmic evolution can give rise to increasing of energy, and it
can be the origin of formation of galaxies and their clusters due
to a capture of cosmic objects by the central gravitational field.
In this paper, we consider the phenomenon  of capture of objects
by gravitational fields in the expanding Universe for any
equations of state and the central fields to compare different
cosmological models. The article starts with the determination of
the gravitational potential as a component of the metrics by the
cosmological perturbation theory (Section 1). In Section 2, cosmic
evolution is described. Section 3 describes the dynamics of a free
particle in the expanding Universe with the
Friedman---Lem\^aitre---Robertson---Walker (FLRW) metrics.
Sections 4 and 5, are devoted to dynamics and capture of a
particle in the Newton--like potential.  The conclusion completes
this paper.

\section{Metrics}
We consider the motion of a test particle in the space with an
interval
 \be\label{ds:psi} ds^2 = a(\eta)^2 \left[
\left( 1-\Phi \right) d\eta^2 - \left( 1+\Phi \right) \left(
dx^i+N^i d\eta \right)^2 \right], \ee where the metric components
can be determined by  the Einstein equations in the cosmological
perturbation theory \cite{Lifshits:63,Kodama:84,Bardeen:80}:
 \be \label{eq:T:00:sigma}
a^2 \Delta \Phi &=& - 8 \pi G T_0^0,\\
\label{eq:T:kk:sigma}\left[\frac{a^2}{2}
\left(2\partial_iN^i-3\Phi'\right)\right]' &=& 8 \pi G T_k^k, \ee
where $\Phi'=d\Phi/d\eta$ and $T_0^0$ and $T_k^k$ are the
components of an energy--momentum tensor of the gravitational
center. Equations (\ref{eq:T:00:sigma}), (\ref{eq:T:kk:sigma})
 show us that there is a reference frame with a zero pressure $T_k^k=0$
 and
nonzero   shift vector $2\partial_iN^i=3\Phi'$ in which  the
solution of equations (\ref{eq:T:00:sigma}), (\ref{eq:T:kk:sigma})
take the Newton-like form \be \Phi(x)= 2 G \int d^3y
\frac{T_{00}(y)}{|y-x|}, ~~~~~~~~~~~~~N^i(x)=\partial^i\int
d^3y\frac{3}{2}\frac{\Phi'(y)}{{|y-x|}}\ee in the case of an
arbitrary equation of state. For a point source
$T_{00}=M\delta^3(x)$ it takes the form $\Phi_{\rm point} = {r_g
}/{r}$, where $r_g=2GM$. The cylindric symmetry of a source $
T_{00}=\frac{M}{2l} \delta(x_1) \delta(x_2) \theta(l-|x_3|) $
leads to another potential with the derivative
\be\label{eq:Phi:cyl} \frac{d\Phi_{\rm cyl.}}{dr}
=-\frac{r_g}{r\sqrt{l^2+r^2}}\ee instead of $ {d\Phi_{\rm
point}}/{dr} = -{r_g}/{r^2}$.

\section{Cosmic evolution} The dependence of
the scale factor ($a$) on the conformal time ($\eta$) is given by
the Einstein---Friedmann equation \cite{Narlikar}:
\be\label{a:Omega}
H_0^{-1}\left(\frac{da}{d\eta}\right)^2=\Omega(a)\equiv\Omega_{\rm
Stiff}a^{-2} +\Omega_{\rm Radiation}+\Omega_{\rm
Matter}a+\Omega_{\Lambda}a^4\ee where $\Omega_i$ is the sum of the
partial densities: stiff, radiation, matter, and
$\Lambda$--term--state, respectively, normalized, by the unit
($\Omega\Big{|}_{a=1}=1$); $H_0$ is the present--day value of the
Hubble parameter. The best fit to the Supernova data \cite{snov}
requires a cosmological constant $\Omega_{\rm Stiff}=0$,
$\Omega_\Lambda=0.7$ and $\Omega_{\rm Matter}=0.3$ in the case of
Standard cosmology (SC), where the measurable distance is
identified with the world space interval \be\label{R:r} R^C{\rm
measurable}=R=a(t)r;\quad r=(x^1)^2+(x^2)^2+(x^3)^2. \ee In the
conformal cosmology \cite{Hoyl-Narlikar,039}, measurable time and
distance are identified with the conformal quantities $(r,\eta)$
\be R^C{\rm measurable}=r;\quad d\eta=dt/a(t). \ee In this case
the Supernova data \cite{snov} are consistent with the dominance
of the stiff state \cite{039,Danilo}, $\Omega_{\rm Stiff}\simeq
0.85 \pm 0.15$, $\Omega_{\rm Matter}=0.15 \pm 0.15$. In the case
$\Omega_{\rm Stiff}=1$, we have the square root dependence of the
scale factor on conformal time \be
a(\eta)=\sqrt{1+2H_0(\eta-\eta_0)}\ee that does not contradict the
standard description of nucleosynthesis \cite{Danilo,three} where
the Friedmann time is replaced by the conformal one.

\newpage

\section{``Free'' motion of a particle in FRWL metrics}
A free motion of a particle in the conformal--flat metrics
 \be\label{ds:FRWL}
(ds^2)=(dt)^2-a^2(t)(dx^i)^2=a^2(\eta)\left[d\eta^2-(dx^i)^2\right]
\ee follows form the definition of the corresponding one--particle
energy in the field theory \cite{Grib} \be E=p_0
=\sqrt{p_i^2+m_0^2a^2(\eta)}=m_0a(\eta)+\frac{p^2}{2m_0a(\eta)}.\nonumber\ee
The nonrelativistic action \be\label{S:eta}
S_0=\int\limits_{\eta_I}^{\eta_0} d\eta \left[ p_i
x'_i-p_0+m_0a(\eta) \right] \simeq
\int\limits_{\eta_I}^{\eta_0}d\eta\left[p_i x'_i -
\frac{p_i^2}{2m_0a(\eta)} \right] \ee coincides, in terms of the
``world time'' $t$, with the action considered in \cite{Peebles}
\be\label{S:Peebles} S_0=\int\limits_{t_I}^{t_0}dt\left[p_i \dot
x_i - \frac{p^2}{2m_0a^2(t)} \right]\quad \left(\dot x =
\frac{dx_i}{dt}\right).\ee All solutions of the equations derived
from (\ref{S:Peebles}) can be obtained by the conformal
transformation of the solutions \be\label{sol:free:x} x_i(\eta)
=x_i^{(0)}+\frac{p_i^{(0)}}{m_0}\Int{\eta_0}{\eta}\frac{d\eta}{a(\eta)}
=x_i^{(I)}+\frac{p_i^{(I)}}{m_0}\Int{\eta_I}{\eta}\frac{d\eta}{a_I(\eta)}\ee
of the equations derived from the action (\ref{S:eta}), where
$a_I= a(\eta=\eta_I)$, $p_i^{(I)}$ and $x_i^{(I)}$ can be
arbitrary initial data at the moment $\eta=\eta_I$.

Physical results do not depend on the choice of variables in the
action (\ref{S:eta}), if observable quantities are defined
unambiguously. But physical results depend on the choice of
observables (as we have seen above in Section 1).

\section{Newtonian motion of particles in the expanding Universe}
The energy of a particle (moving on the geodesic line in space
with set metrics) can be found by solving the mass--shell
equation. Consider metrics (\ref{ds:psi}) that denote
$g_{\mu\nu}$. Calculating the integral of motion of a test
particle in this metric: \be p^2=g^{\mu\nu} p_\mu p_\nu = m^2, \ee
we can find an expression for energy $p_0$ \be\label{eq:approx}
p_0 \approx -N^r p_r+
\pm\left(1-\frac{r_g}{2r}\right)\left[\left(1-\frac{r_g}{2r}\right)m
+ \frac{p_r^2}{2m} +\frac{p_\theta^2}{2mr^2} \right], \ee where
$N^r=- \frac 34 r_gH $ is the radial component of the shift
vector. For a positive sign we get the action
\be\label{S:classic}\label{S:eta:alpha} S &=&
\int\limits_{\eta_I}^{\eta_0} d\eta \left[p_r r' +p_\theta \theta'
- E\right], \ee where $E=p_0-m$. In the nonrelativistic limit of
small velocities the energy takes the form \be\label{E:classic} E
\approx E_{\rm
classic}=\frac{p_r^2}{2m}+\frac{p_\theta^2}{2mr^2}-\frac{r_gm}{2r},\ee
here $m=m_0 a(\eta)$ is the mass a test particle that depends on
time. The product $r_gm$ is a conformal invariant and does not
depend on time.

The action (\ref{S:eta:alpha}) shows us a possibility of capturing
in a particle with the running mass $m(\eta)=m_0a(\eta)$ at the
moment when the kinetic term $\sim 1/a$ takes a value equal to the
potential ($\alpha/r$). To study this effect of capture, we pass
to the scale factor $a(\eta)$ as the time--like parameter, using
the cosmological equations (\ref{a:Omega}), dimensionless
variables $y=r/r_0$, $p_y=p_r/m_0$, and three velocities: orbital
($v_0$), Newtonian ($w_0$), and cosmic ($c_0$) defined as
$v_0={P_\theta}/{(r_0m_0)}$, $w_0=\sqrt{\alpha/{(r_0m_0)}}$, and
$c_0=H_0r_0.$

In this case, the equation derived from (\ref{S:eta:alpha}) takes
the form  \be\label{eq:v:w} \frac{v_0^2}{y^3}=\frac{w_0^2a}{y^2}
+c_0^2(\Omega^{1/2}(a)a)\frac{d}{da}
\left[(\Omega^{1/2}(a)a)\frac{dy}{da}\right]. \ee The general
solution of their equation is given in \cite{astro-ph:0206114} for
the stiff state $\Omega^{1/2}(a)a$ $=1$. This solution allows us
to check the results of the numerical calculation of eq.
(\ref{eq:v:w}) for arbitrary parameters $\Omega_{\rm Stiff}$,
$\Omega_{\rm Radiation}$, $\Omega_{\rm Matter}$, and
$\Omega_\Lambda$.

For the case $c_0=0$, we get the class of circular trajectories
\be\label{y:a} y(a)=\frac{y(0)}{a}; \quad y(a)=1\ee with the
constraint of the initial velocities $v_0^2=w_0^2$ well known as
the ``virial theorem'' \cite{Chernin}:
$v_0=\sqrt{{r_g}/{(2r_0)}}$. In the case of the cylindric symmetry
of a sources (\ref{eq:Phi:cyl}) we get
$v_0=\sqrt{r_g/\left(2\sqrt{l^2+r_0^2}\right)}$. The last formula
can explain the constant rotational curves in the region $r_0
\lesssim l$.

To estimate the role of the last term in eq. (\ref{eq:v:w}) in the
class of circular trajectories, we  substitute (\ref{y:a}) in eq.
(\ref{eq:v:w}) for $c_0^2\neq 0$ and $\Omega=\Omega_{\rm Stiff}
a^{-2}+\Omega_{\rm Matter}a + \Omega_\Lambda a^4$. In this way, we
get the following relation between the velocities $v_0$, $w_0$,
and $c_0$: \be v_0^2=w_0^2+c_0^2\left[2-\left(\frac 32 \Omega_{\rm
Matter}+ 3 \Omega_\Lambda\right)\right].\ee In the case of the
Conformal Cosmology (CC), $\Omega_{\rm Matter}=\Omega_\Lambda=0$,
these equations become \be v_0^2=w_0^2+2c_0^2, \ee and the last
term $2c_0^2$ plays the role of the Dark Matter, whereas in the
case of Standard Cosmology (SC): $\Omega_{\rm Matter}=0.3$,
$\Omega_\Lambda=0.7$, the last term is negative so that
$v_0^2=w_0^2-0.5 c_0^2$. Thus the standard cosmology requires one
more Dark Matter \cite{astro-ph:0301543} in contrast to the
conformal cosmology \cite{039}.

\section{Capture of cosmic objects by the central field}
Let us consider the action (\ref{S:eta:alpha})--(\ref{E:classic})
in terms of the Friedmann coordinates $R=ra(\eta)$,
$P_R=p/a(\eta)$, and $dt=a(\eta)d\eta$:
 \be\label{cr11ar}
 S_A=\int\limits_{t_I}^{t_0}dt\left[P_R(\dot R-HR)+P_\theta \dot \theta-
 \frac{P_R^2}{2m_0} -\frac{P_\theta^2}{2m_0R^2}+\frac{\alpha}{R}\right].
 \ee

The ``energy'' of a particle in (\ref{cr11ar}) \be \label{E}
E=HRP_R+\frac{P_R^2}{2m_0}+\frac{P_\theta^2}{2m_0R^2}-
 \frac{\alpha}{R}
\ee is not conserved in contrast to the        energy of a
particle with a constant mass in the Newtonian  mechanics. This
energy can change sign at $t=t_I$.  It is known that the change of
a sign of the energy means the change of an unrestricted motion of
a particle by a finite motion in the central field. Therefore, the
time $t_0$ can be treated as the time of the capture of a particle
(cosmic object) by the central gravitational field
\cite{astro-ph:0301543}.

\begin{figure}[htb!]
\begin{center}
\includegraphics[width=0.6\textwidth]{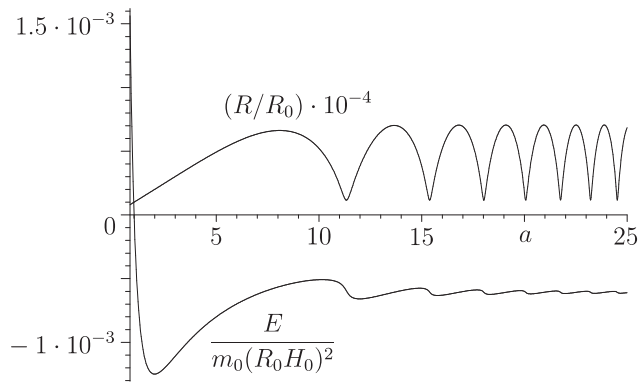}
\end{center}
\caption{\small Effect of capture of a cosmic object by the
central field is described by the dynamics of a radial coordinate
$R=ar$ and its energy $E$ (\ref{E}).} \label{fig:ER}
\end{figure}

For the local Group $M_{LC}=10^{13}\div 10^{14}$, $R_{LC}=8~{\rm
Mpc}$, $c_0\sim 500 ~{\rm km/s}$, and  $v_0\sim 100  ~{\rm
km/s}\div 200 ~{\rm km/s}$. The class of ellipsoidal trajectories
of capture explains the anisotropic local velocity field of nearby
galaxies by their Newtonian motion \cite{Karach,astro-ph:0206114}.

\section{Conclusion}
It is shown that the cosmic evolution can be the mechanism of a
capture of cosmic objects that can form halos of galaxies and
their clusters. This capture can replay one of the fundamental
questions of modern theory of formation of galaxies: How is a
system of unbounded ``particles'' with positive energy converted
to the system of bounded ``particles'' with negative energy? The
range of validity of the Newtonian mechanics for description of
galaxies is restricted by the critical radius $R_{cr}\simeq
10^{20}$ cm $(M/M_\odot)^{1/3}$ that is closed to the radius of
surface of the zero--velocity, separating a cluster from the
cosmological evolution.  We managed to express the Newton-like
equation  in terms of the red shift parameter for an arbitrary
equations of state and to  obtain modification of the rotational
curves. It  shows us that the cosmological evolution can plays the
role  ascribed to the Dark Matter.
 The deficit of the visible matter can be
increased in the version of  cosmology where observational
quantities are identified with the conformal ones; whereas
 the standard cosmological model requires one more Dark Matter.

The authors thank B.M. Barbashov, D. Blaschke, D.V. Gal'tsov, and
 V.A. Rubakov for usefull discussions.

\end{document}